\documentclass[final,1p,times]{elsarticle}

\usepackage{amssymb}

\journal{Nuclear Physics A}

\biboptions{sort&compress}

\begin{document}

\begin{frontmatter}


 \author[dfg]{M. Maggiora\corref{distocoll}}
 \ead{marco.maggiora@to.infn.it}
 \cortext[distocoll]{on behalf of the DISTO collaboration}
\address[dfg]{Dipartimento di Fisica Generale ``A. Avogadro'' and INFN, Torino, Italy}

\title{DISTO data on $K^-pp$}


\author[smi,ecu]{P. Kienle}
\author[smi]{K. Suzuki}
\author[tokyo,riken]{T. Yamazaki}
\author[dfg]{M. Alexeev}
\author[dfg]{A. Amoroso}
\author[dfg]{F. Balestra}
\author[saclay]{Y. Bedfer\fnref{bedfer}}
\author[dfg,saclay]{R. Bertini}
\author[iucf]{L. C. Bland\fnref{bland}}
\author[giessen]{A. Brenschede\fnref{arndt}}
\author[saclay]{F. Brochard}
\author[dfg]{M. P. Bussa}
\author[saclay]{S. Choi\fnref{choi}}
\author[dfg]{M. L. Colantoni}
\author[rossen]{R. Dressler\fnref{dressler}}
\author[iucf]{M. Dzemidzic\fnref{mario}}
\author[saclay]{J. Cl. Faivre}
\author[dfg]{L. Ferrero}
\author[kra1,kra2]{J. Foryciarz\fnref{jurek}}
\author[giessen]{I. Fr\"ohlich\fnref{ingo}}
\author[jinr]{V. Frolov\fnref{frolov}}
\author[dfg]{R. Garfagnini}
\author[dfg]{A. Grasso}
\author[dfg,saclay]{S.~Heinz\fnref{heinz}}
\author[iucf]{W. W. Jacobs}
\author[giessen]{W. K\"uhn}
\author[dfg]{A. Maggiora}
\author[po]{D. Panzieri}
\author[dfg]{S. Sosio}
\author[giessen]{H. W. Pfaff}
\author[dcfg,jinr]{G. Pontecorvo}
\author[jinr]{A. Popov}
\author[giessen]{J. Ritman\fnref{jim}}
\author[kra1]{P. Salabura}
\author[jinr]{V. Tchalyshev}
\author[iucf]{S. E. Vigdor\fnref{steve}}

\fntext[bedfer]{Present address: DAPNIA/SPhN, CEA Saclay, F}
\fntext[bland]{{Present address: BNL, USA}}
\fntext[arndt]{Present address: DIAMOS AG, Sulzbach, D}
\fntext[choi]{Present address: Seoul National University, Seoul, KR}
\fntext[dressler]{Present address: Paul Scherrer Institut, Villigen, CH}
\fntext[mario]{Present address: IU School of Medicine, Indianapolis, USA}
\fntext[jurek]{Present address: Motorola Polska Software Center, Krak\'ow, PL}
\fntext[ingo]{Present address: IKF, Frankfurt, D}
\fntext[frolov]{Present address: Dip. di Fisica Generale and INFN, Torino, I}
\fntext[heinz]{Present address: GSI, Darmstadt, D}
\fntext[jim]{Present address: FZ, Juelich, D}
\fntext[steve]{Present address: BNL, USA}

\address[smi]{Stefan Meyer Institute for Subatomic Physics, Austrian Academy of Sciences, Vienna, Austria}
\address[ecu]{Excellence Cluster Universe, Technische Universit ̈t M ̈nchen, Garching, Germany}
\address[tokyo]{Department of Physics, University of Tokyo, Tokyo, 116-0033 Japan}
\address[riken]{RIKEN Nishina Center, Wako, Saitama, 351-0198 Japan}
\address[saclay]{Laboratoire National Saturne, CEA Saclay, France}
\address[iucf]{Indiana University Cyclotron Facility, Bloomington, Indiana, U.S.A.}
\address[giessen]{II. Physikalisches Institut, Univ. Gie\ss{}en, Germany}
\address[jinr]{JINR, Dubna, Russia}
\address[kra1]{M.~Smoluchowski Institute of Physics, Jagellonian University, Krak\'{o}w, Poland}
\address[kra2]{H.~Niewodniczanski Institute of Nuclear Physics,Krak\'{o}w, Poland}
\address[po]{Universit\`{a} del Piemonte Orientale and INFN, Torino, Italy}
\address[rossen]{Forschungszentrum Rossendorf, Germany}

\begin{abstract}
The data from the DISTO Collaboration on the exclusive $pp \rightarrow p  K^+  \Lambda$ production acquired at $T_p=2.85~GeV$ have been re-analysed in order to search for a deeply bound $K^- pp (\equiv X)$ state, to be formed in the binary process $pp \rightarrow K^+ X$. The {\it preliminary} spectra of the $\Delta M_{K^+}$ missing-mass and of the $M_{p\Lambda}$ invariant-mass show, for large transverse-momenta of protons and kaons, a distinct broad peak with a mass $M_X = 2265 \pm 2~ MeV/c^2$ and a width $\Gamma_X = 118 \pm 8 ~MeV/c^2$.
\end{abstract}

\begin{keyword}
$\bar K$ nuclei \sep kaon condensation \sep super-strong nuclear force \sep strange
di-baryon 
\PACS

\end{keyword}

\end{frontmatter}



\begin{table}[b!]
    \centering
    \begin{tabular}{|c|c|c|}
    \hline
&&\\[-3mm]
\large Reaction & \large $T_{p,thr}~(GeV)$ & \large 
Charged particle trigger\\
&&\\[-4mm]
    \hline
$\vec{p}p\rightarrow p K^+ \vec{\Lambda}$ & 1.58 & $pK^+(p \pi^-)$\\
    \hline
&&\\[-3mm]
$\vec{p}p\rightarrow p K^+ \vec{\Sigma}^0 \quad,\quad \vec{\Sigma}^0\rightarrow \vec{\Lambda}\gamma$
& 1.79 & $pK^+(p \pi^-)$\\
&&\\[-3mm]
    \hline
$\vec{p}p\rightarrow p K^+ \Sigma^{*0}_{(1385)}$ & 2.34 & 
\begin{minipage}[h]{.35\textwidth}\vspace{1mm}\centering
$pK^+(p\pi^-)$ from $\Lambda\pi^0$ or $\Sigma^0\pi^0$\\
$pK^+\pi^+(\pi^-)$ from $\Sigma^-\pi^+$\\
$pK^+\pi^-(p)$ or $(\pi^+)$ from $\Sigma^+\pi^-$\\
\vspace{1mm}
\end{minipage}\\
    \hline
$\vec{p}p\rightarrow p K^+ \Lambda^{*}_{(1405)}$ & 2.40 & 
\begin{minipage}[h]{.35\textwidth}\vspace{1mm}\centering
$pK^+\pi^+(\pi^-)$ from $\Sigma^-\pi^+$\\
$pK^+(p\pi^-)$ from $\Sigma^0\pi^0$\\
$pK^+\pi^-(p)$ or $(\pi^+)$ from $\Sigma^+\pi^-$\\
\vspace{1mm}
\end{minipage}\\
    \hline
    \end{tabular}

    \caption{Exclusive hyperon production channels that can be accessed at $T_p=2.85~GeV$. The particles selected for each single final state by the charged particle trigger, a trigger with charged particle multiplicity four, can be found in the right column, the () evidencing those particles emerging from a displaced secondary vertex .\label{tab:channels}}
   \end{table}

The simplest kaonic nuclear bound system $K^-pp$ is currently interpreted in the literature in two different kinds of scenarios: the {\it strong binding regimes} and the {\it weak binding regimes}.
In the former case the $K^-pp$ has been predicted to be a quasi-stable state with mass $M = 2322~ MeV/c^2$, binding energy $B_K = 48$ MeV and partial decay width $\Gamma_{\Sigma \pi p} = 61$ MeV \cite{Akaishi:2002bg,Yamazaki:2002uh}. The strong binding arises from the migration of the $K^-$ between the two protons in a molecule-like structured $K^-pp$, causing a "super-strong nuclear force" \cite{Yamazaki:2007hj,Yamazaki:2007cs}. Such a picture originates from the ansatz that the $\Lambda_{(1405)}$ resonance is an $I=0$ $\bar{K} N$ quasi-bound state embedded in the $\Sigma \pi$ continuum. 
Recent Faddeev calculations \cite{Shevchenko:2006xy,Shevchenko:2007zz,Ikeda:2007nz} lead as well to a deeply bound $K^-pp$ state.
A different approach based on chiral dynamics \cite{Jido:2003cb,Magas:2005vu,Hyodo:2007jq} places the $K^-p$ pole close to the $\bar{K}N$ threshold, at $1420\div1430~MeV/c^2$, leading to a weak $\bar{K}N$ interaction and to a shallow $K^-pp$ bound state \cite{Dote:2008hw}.

Whereas the dispute on the location of the $K^-p$ state, at $1405$ or at $1420$, remains unresolved, the importance to distinguish between the {\it strong binding} and the {\it weak binding regimes} by an experimental investigation of the $K^-pp$ formation in $pp$ reactions is supported by its implication on kaon condensation \cite{Kaplan:1986yq,Brown:1993yv} and on the existence of dense kaonic nuclear states \cite{Akaishi:2002bg,Yamazaki:2002uh,Dote:2002db,Dote:2003ac,Yamazaki:2003hs}.

The formation of a strongly bound $K^-pp$ system with a short $p$-$p$ distance in a $pp \rightarrow K^+ + K^-pp$ reaction has been predicted \cite{Akaishi:2002bg,Yamazaki:2002uh} as the consequence of a significantly large sticking probability between $\Lambda_{(1405)}$ and $p$, due to the short range and large momentum transfer of the $pp$ reaction \cite{Yamazaki:2007hj,Yamazaki:2007cs}. We report herewith the search for a possible candidate of such an exotic  $K^-pp$ deeply bound state in the  existing experimental data acquired by the DISTO spectrometer for the $p  p \rightarrow p K^+  \Lambda$ reaction.

\section{Data sample}
\label{datasample}

The DISTO spectrometer, described in details elsewhere \cite{Balestra:1999ev}, was aimed to investigate hyperon and meson productions in $pp$ collisions, making use of the transversely polarised proton beam of SATURNE (Saclay, France) and of an unpolarised  liquid hydrogen target. The main goal of the hyperon program was determining different spin observables (polarisation, analysing power, depolarisation transfer) \cite{Balestra:1999br,Maggiora:2001tx} in exclusive hyperon production like $p  p \rightarrow p K^+  \Lambda  $ and  $p  p \rightarrow p K^+  \Sigma^0  $, selecting the different exclusive contributions by a complete kinematic reconstruction of the final state, for three different kinetic energies of the beam: $T_p=2.85$, $2.5$ and $2.145~GeV$. The trigger was set in order to acquire events with at least four emerging charged particles. The data sample chosen to probe the existence of a possible bound $K^-pp$ system is constituted by those events acquired at the higher $T_p$ (in such a case those exclusive channel reported in Tab. \ref{tab:channels} can be accessed) for the $pK^+\Lambda$ final state, in which a complete reconstruction of the final state is performed, and no particle goes undetected.



\begin{figure}
	\parbox[c]{\textwidth}{
	  \vspace{-4mm}
  	  \centering
	  \includegraphics[width=.46\textwidth]{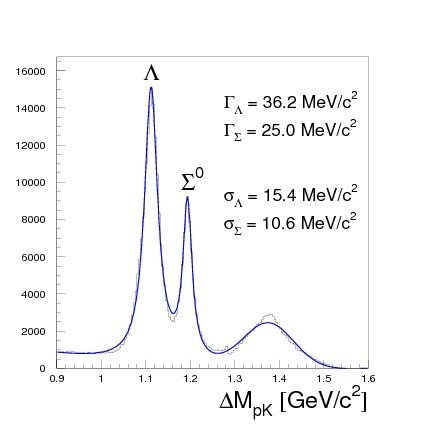}
	  \includegraphics[width=.49\textwidth]{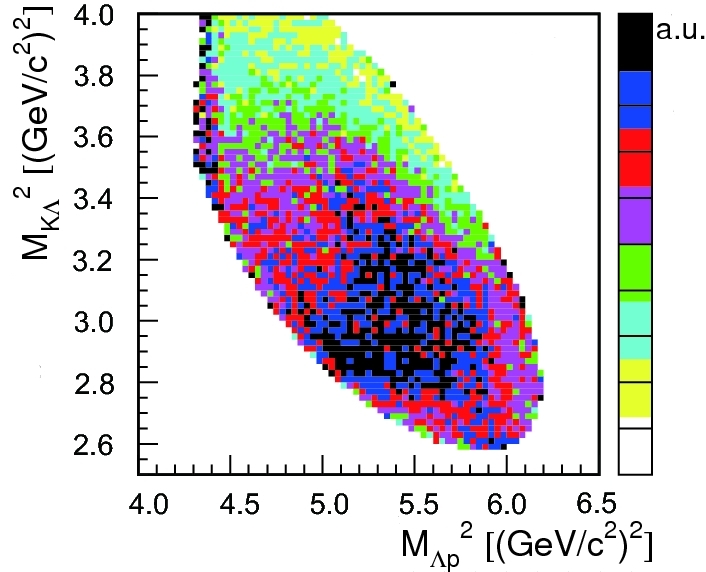}
	  \vspace{-2mm}
	  \caption{All the spectra reported herewith refer {\it only} to those events that survived the kinematically constrained refit described in the text. On the left, the $\Delta M_{pK}$ missing mass spectrum post-refit ($\Delta M_{pK}$ is left free in the refitting procedure). On the right, the acceptance corrected Dalitz plot, $M(\Lambda p)^2$ vs $M(K\Lambda)^2$, of $pp \rightarrow p K^+ \Lambda$ at $T_p=2.85~GeV$.}
\label{fig:mm_dalitz}
	}
\end{figure}

A first event selection is performed asking for: a $\pi^+$-veto (no track in the final state positively identified as a $\pi^+$); the positive track emerging from the displaced secondary vertex to be clearly identified as a proton (the proton of the weak-decay $\Lambda\rightarrow p \pi^-$); its polar angle in the $\Lambda$ rest frame to be bounded to its kinematic limit ($|\theta_p^{CM_\Lambda}|<0.15~rad$); a primary vertex reconstructed within the target region ($|z_{V_R}|<3.5~cm$); a minimum $\Lambda$ decay length of $4~cm$. The events in which a $\Lambda$ has been produced are identified through the invariant-mass of the two tracks emerging from the displaced secondary vertex ($M_{p\pi^-}\sim M_\Lambda$), and then fed to a refit constrained by the following kinematic requirements: the reconstructed invariant mass $M_{p\pi^-}$ is constrained to the $\Lambda$ mass ($M_{p\pi^-}=M_\Lambda$); the momentum conservation at the secondary vertex is enforced by constraining the reconstructed $\Lambda$ momentum along the vertex joiner direction ($\vec{p}_\Lambda \parallel \overrightarrow{V_RV_D}$); the four-body missing mass is constrained either to 0 ($\Delta M_{4b}= 0$ selects reactions 1 and 2 in Tab. \ref{tab:channels}) or to the pion mass ($\Delta M_{4b}= M_\pi$ selects reactions 3 and 4 in the same table). A further "soft" constraint requiring the reaction vertex to be along the beam line is achieved including in the event refitting $~\chi^2$ a term esplicitly accounting for the distance between the reaction vertex and the beam line. A tight requirement on the event refitting $~\chi^2$ is the most effective cut to reach background rejection and hyperon separation much better than those achieved in earlier stages of these data analyses \cite{Balestra:1999br}. 
 
We can finally resolve the directly produced $\Lambda$'s from those coming from the decays of heavier hyperons by the mean of the $\Delta M_{pK}$ missing-mass spectrum shown in Fig. \ref{fig:mm_dalitz}.a: the lower mass peak correspond to the exclusive $pK^+\Lambda$ final state; the second peak to reaction 2 of Tab. \ref{tab:channels}, while the bump at higher mass values may contain contributions arising from reactions 3 and 4. The data sample corresponding to the exclusive reaction channel:
\begin{equation}
p  p \rightarrow p K^+ \Lambda
\label{eq:pp2pKL}
\end{equation}
that can be extracted from the DISTO data acquired 
at a beam kinetic energy $T_p=2.85~GeV$, is composed of $\sim177~K$ events, which fullfill all the requirements described above, and in particular the tight $~\chi^2$ cut in the refit procedure. In these kinematic conditions a candidate of the $K^-pp$  bound state ($\equiv X$) could be formed in the two-body process:
\begin{equation}
p  p \rightarrow K^+  X \quad,\quad X \rightarrow p + \Lambda
\label{eq:pp2KX}
\end{equation}
and contribute to the low mass peak of Fig. \ref{fig:mm_dalitz}.a through the $X \rightarrow p  \Lambda$ decay, other processes like free $\Lambda_{(1405)}$ emission being excluded.
The whole $pK^+\Lambda$ sample could then include besides the ordinary three-body process (\ref{eq:pp2pKL}), acting in the present analysis as a "background", also the exotic process of the two-body reaction (\ref{eq:pp2KX}).

A sample of events simulated for the $pK^+\Lambda$ final state according to an uniform phase-space distribution, folded with the DISTO acceptance, and then fed to the complete reconstruction and analysis chain, fulfilling hence the cuts and the refitting procedure described above, leads to the distributions reported herewith and marked {\it SIM}; such distributions have to be compared with the uncorrected experimental data {\it UNC}. In order to minimise the effects of possible uncertainties arising from the evaluation of an efficiency matrix we adopt herewith the {\it deviation spectra} technique, an efficiency-compensated presentation of the experimental data in which the ratio:
\begin{equation}
 DEV = \frac{UNC}{SIM} 
\label{eq:dev}
\end{equation}
is evaluated bin by bin. A {\it DEV} spectrum is different from its corresponding intensity distribution, the latter being affected by phase-space limitation and hence bell-shaped. The {\it DEV} spectrum is free, at first order, both from uncertainties in the efficiency-matrix evaluation and from phase-space constraints. A {\it DEV} 
distribution deviates from its generally flat nature, indicating a structure, when a physically meaningful deviation from the uniform phase-space distribution occurs; that should make the comparison with theoretical predictions easier. In those particular cases in which the physical distribution is expected to be flat (as for example in the case of the Dalitz distribution), and only in those cases, the {\it SIM} spectra reflect just the effects of the acceptance and of the efficiencies, and the {\it DEV} spectra are equivalent to their corresponding efficiency-corrected intensity distributions {\it COR}.

Fig. \ref{fig:mm_dalitz}.b show the acceptance-corrected Dalitz plot
of the considered $pK^+\Lambda$ sample in the plane defined by the two invariant-masses $M_{\Lambda p}^2$ and $M_{K^+\Lambda }^2$. The Dalitz distribution of the
ordinary three-body process (\ref{eq:pp2pKL}) should be different from that corresponding to an uniform phase-space distribution, but is expected to be continuous without any local bump structure \cite{Akapriv:2008}. The Dalitz plot alone cannot resolve the "ordinary" process (\ref{eq:pp2pKL}) from the "exotic" one (\ref{eq:pp2KX}), and yet the plot in Fig. \ref{fig:mm_dalitz}.b reveals some structure hard to be explained by the latter process only. The angular correlations of the three particles in the final state are hence needed to discriminate among these two processes.

\section{Angular correlations}
\label{angcor}

\begin{figure}[t]
		\begin{minipage}[h]{\textwidth}
		\hspace{-3mm}
		\includegraphics[width=.52\textwidth]{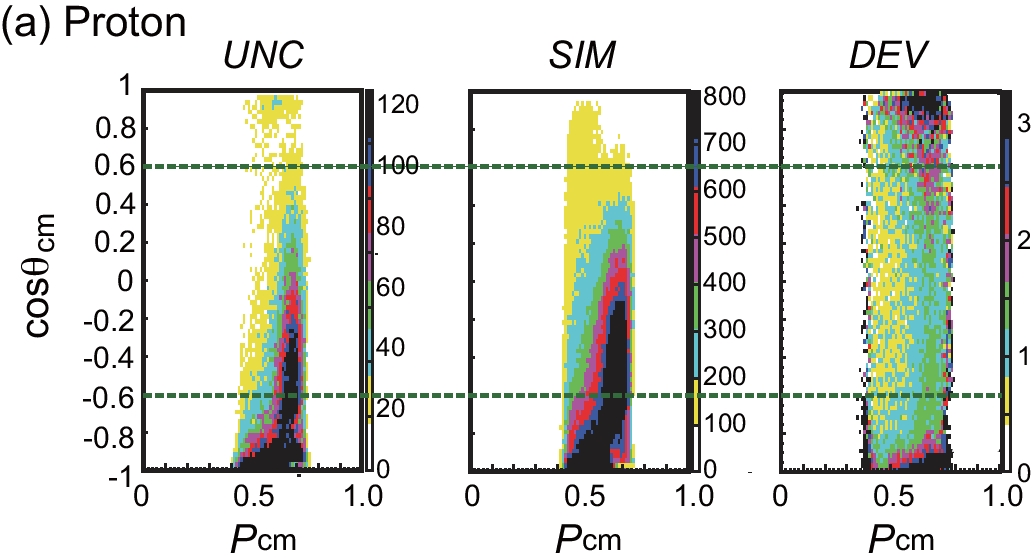}
		\hspace{-2mm}
		\includegraphics[width=.52\textwidth]{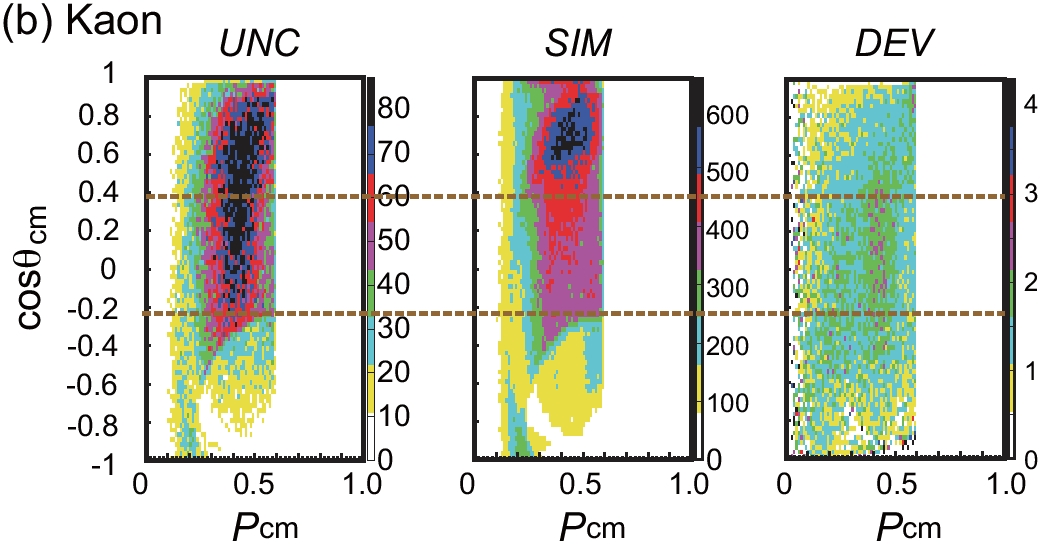}
 	        \vspace{-6mm}
		\end{minipage}
\caption{Uncorrected ({\it UNC}), simulated ({\it SIM}) and deviation ({\it DEV}) distributions, for (a) $p$ and (b) $K^+$, of the momentum $p_{CM}$ versus the polar angle $cos(\theta_{CM}$) in the $pp$ center of mass frame. Dotted lines define the proton-angle  and the kaon-angle cuts, respectively $|cos(\theta_{CM,p})| < 0.6$ and $-0.2 < cos(\theta_{CM,K^+}) < 0.4$.}
\label{fig:ang_cor}
\end{figure}

Fig. \ref{fig:ang_cor} shows the {\it UNC}, {\it SIM} and {\it DEV} distributions of the momentum  $p_{CM}$ versus the polar angle $cos(\theta_{CM}$)  for the proton and the kaon emerging from the primary reaction vertex, respectively. Both the {\it UNC} and the {\it SIM} $p$ distributions are strongly peaked in the backward direction, reflecting the corresponding larger acceptance of the DISTO spectrometer for those events showing a $\Lambda \rightarrow p \pi^-$ decay in the forward direction. An indication of the effectiveness of the deviation spectra technique comes from considering the {\it DEV} distributions for $p$ and $\Lambda$ (here not shown): both are remarkably symmetric and peaked at $cos(\theta_{CM})\pm1$, as expected considering the symmetric $p$-$p$ scattering in the CM frame. 

The dominant contribution from protons at extremely backward angles ($cos \theta_{CM,p} \sim - 0.9$ in Fig. \ref{fig:ang_cor}.a) can be interpreted as composed of low transverse-momentum ($q_T < 0.3~ GeV/c$) events from the "ordinary" process (\ref{eq:pp2pKL}), since the maximum proton momentum in (\ref{eq:pp2pKL}) is $0.751~GeV/c$. The $p$ angular  distribution with respect to the incident beam has been fitted to a theoretical model \cite{Akapriv:2008}, predicting a $p$-$p$ collision length for this dominant component of about $0.9~fm$ (corresponding to an intermediate boson mass of $m_B \sim 0.2~GeV/c^2$).
On the other hand the events characterised by a large proton polar angle can be large-$q_T$ "ordinary" process events, but also events from the "exotic" process (\ref{eq:pp2KX}) involving the decay of $X$ with a transverse-momentum $\sim 0.5~GeV/c$. A cut on the proton-emission polar angle, asking for $|cos(\theta_{CM,p})| < 0.6$, can thus enhance the contribution from the process (\ref{eq:pp2KX}) by rejecting those events with a small proton polar angle, dominated by the process (\ref{eq:pp2pKL}).

The {\it DEV} kaon distribution of Fig. \ref{fig:ang_cor}.b shows a clear mono-energetic component for $p_{CM,K^+}\sim0.4~GeV/c$, a possible signature from a two-body process (\ref{eq:pp2KX}). It should be stressed that: such a component is evident even before enhancing the contribution from reaction (\ref{eq:pp2KX}) by applying the proton-angle cut described above; this structure
is already present in the {\it UNC} distribution, and cannot be a fake introduced by the deviation spectra technique, since the {\it SIM} distribution is smooth in the corresponding region.

\begin{figure}[t!]
	\centering
		\includegraphics[width=.49\textwidth]{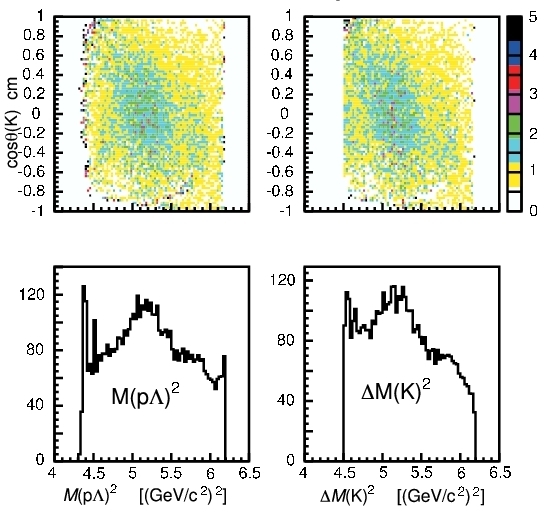}
		\includegraphics[width=.49\textwidth]{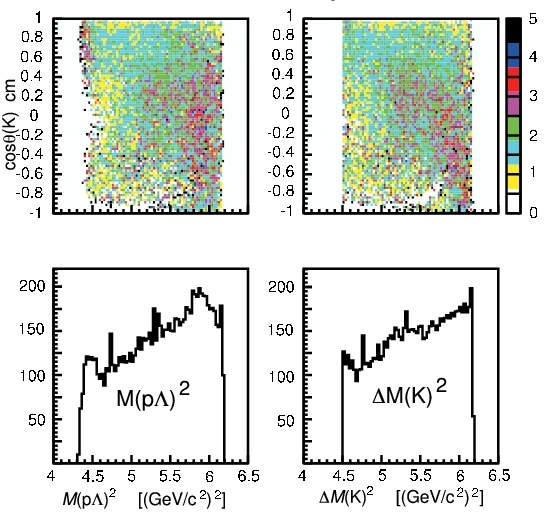}
	\vspace{-2mm}
	\caption{{\it DEV} spectra, for large (left four frames) and small (right four frames) proton-emission polar angle in the CM frame, of the invariant mass $M_{p\Lambda}^2$ and of the missing mass $\Delta M_{K^+}^2$ (bottom frames), and of their correlation with the kaon polar angle in the CM frame $cos(\theta_{CM,K^+})$ (top frames).}
\label{fig:sig_bck}
\end{figure}

\begin{figure}[t!]
	\begin{center}
	    	\includegraphics[width=.49\textwidth]{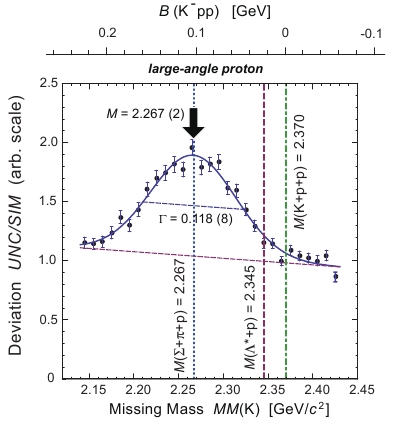}
		\includegraphics[width=.50\textwidth]{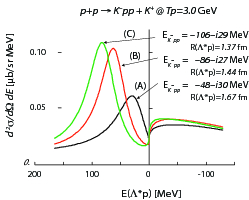}
		\parbox[c]{\textwidth}{
		  \vspace{3mm}\hspace{-3.5mm}
		  \parbox[c]{.50\textwidth}{
		    \centering\includegraphics[width=.50\textwidth]{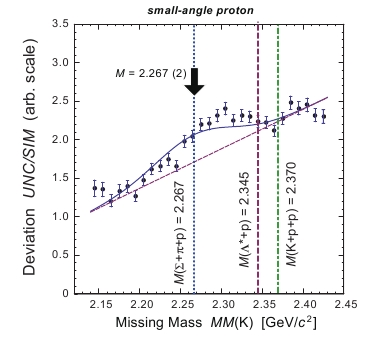}
		  }
		  \parbox[c]{.49\textwidth}{
			\caption{Left frames: {\it DEV} spectra of the $\Delta M_{K^+}$ missing-mass for (top) large-$q_T$ protons and kaons ($|cos(\theta_{CM,p})| < 0.6$, $-0.2 < cos(\theta_{CM,K^+}) < 0.4$) and for (bottom) large-$q_T$ kaons and small-$q_T$ protons ($|cos(\theta_{CM,p})| > 0.6$, $-0.2 < cos(\theta_{CM,K^+}) < 0.4$). The mass spectra are fitted by a Gaussian plus a linear background, and compared with the most relevant particle-emission thresholds; the observed structure is located close to the $\Sigma\pi$ threshold. Right frame: predicted \cite{Yamazaki:2007cs} cross sections of the $pp\rightarrow K^+ + K^-pp$ reaction at $T_p=3~GeV$ as a function of $E_{\Lambda_{(1405)}p}$ for different rms distances $R(\Lambda_{(1405)}p)$ in the case of: A) the original Akaishi-Yamazaki interaction ($B_K= 48~MeV$) \cite{Akaishi:2002bg}; B) a $17\%$ enhancement ($B_K= 86~MeV$) and C) a $25\%$ enhancement ($B_K= 106~MeV$). }
			\label{fig:fits}
		  }
		}
	\end{center}
\end{figure}

To probe the effectiveness of the proton-emission angle cut we can consider in Fig. \ref{fig:sig_bck} the {\it DEV} spectra, for large-angle (left) and small-angle (right) protons, of the invariant mass $M_{p\Lambda}^2$ and of the missing mass $\Delta M_{K^+}^2$ (bottom frames), and their correlation with the kaon polar angle in the CM frame $cos(\theta_{CM,K^+})$ (top frames)\footnote{The uni-dimensional {\it DEV} $M_{p\Lambda}^2$ and $\Delta M_{K^+}^2$ distributions have been obtained performing the ratio of the corresponding uni-dimensional {\it UNC} and {\it SIM} distributions.}. 
The vertical structure in the $p_{CM,K^+}$ spectra of Fig. \ref{fig:ang_cor}.b shows up at the corresponding $\Delta M_{K^+}^2$ values, and is clearly enhanced by asking for large proton-emission polar angles, for which both the {\it DEV} spectra of $M_{p\Lambda}^2$ and $\Delta M_{K^+}^2$ show a structure around the abscissa $x \sim 5.15$, that would correspond to a mass for a $K^-pp$ candidate of $M_X \simeq 2.27~GeV/c^2$. On the other hand the {\it DEV} unidimensional spectra for small proton-emission polar angle (right frames of Fig. \ref{fig:sig_bck}) do not show such a structure; their linear shapes with constant gradients are well accounted for by the "ordinary" process (\ref{eq:pp2pKL}) assuming a collision length  $\hbar/m_B c$ with $m_B \sim 0.2~GeV/c^2$ \cite{Akapriv:2008}. The "ordinary" process (\ref{eq:pp2pKL}) could contribute to the large
proton-angle events as well, but with a shorter collision length, and leading to a flat distribution without gradient \cite{Akapriv:2008}, in great contrast with the  observed {\it DEV} distributions, far from a flat shape alone and showing the large structure on a flat background.

The mono-energetic band in the kaon distribution is clearly enhanced for large $K^+$-emission polar angles, as can be seen in the {\it DEV} of spectrum of Fig. \ref{fig:ang_cor}.b, a spectrum obtained performing no cut on the proton polar angle; the $cos(\theta_{CM,K^+}) < -0.2$ region is depleted
in both the {\it UNC} and {\it SIM} spectra. A further cut on the $K^+$ CM polar angle, $-0.2 < cos(\theta_{CM,K^+}) < 0.4$,  can hence be defined  in order to enhance the possible two-body signature in the considered $pK^+\Lambda$ sample. The final {\it DEV} spectra will thus be obtained asking both for large proton- and large $K^+$-emission polar angles (i.e. large transverse momenta).

In the present preliminary analysis we have shown how a large structure can be observed in the {\it DEV} spectra of both the $M_{p\Lambda}$ invariant-mass and the $\Delta M_{K^+}$ missing-mass (Fig. \ref{fig:sig_bck}); such a structure is enhanced selecting separately either large proton-emission or large kaon-emission polar angles in the CM frame, and such enhancement is maximum under the combined effect of both cuts. The large broad peak that appears in the left top frame of Fig. \ref{fig:fits}, showing the {\it DEV} spectra of the $\Delta M_{K^+}$ missing-mass for large transverse-momenta protons and kaons ($|cos(\theta_{CM,p})| < 0.6$, $-0.2 < cos(\theta_{CM,K^+}) < 0.4$), can be compared with the much less pronounced peak of the left bottom frame of Fig. \ref{fig:fits}, obtained considering the same {\it DEV} spectra for large transverse-momenta kaons and small transverse-momenta protons ($|cos(\theta_{CM,p})| > 0.6$, $-0.2 < cos(\theta_{CM,K^+}) < 0.4$). If we interpret the observed structure as the signature of a two-body process like  (\ref{eq:pp2KX}), a simple fit, by a Gaussian super-imposed over a linear background, of the top-left {\it DEV} spectrum of Fig.  \ref{fig:fits} leads to: 
\begin{equation}
M_X = (2.265 \pm 0.002) ~~GeV/c^2\quad\quad
\Gamma_X  = (0.118 \pm 0.008) ~~GeV/c^2
\label{eq:X}
\end{equation}
The huge difference between the reduced $~\chi^2$'s obtained considering a Gaussian plus a linear background ($~\chi^2/ndf = 34.2/24=1.4$) and the linear background only ($~\chi^2/ndf = 947/27 \sim 35$) provides a statistical significance of about $26~\sigma$.

\section{Concluding remarks}
\label{concl}

The observed structure in the top left frame of Fig. \ref{fig:fits} could be interpreted as a possible candidate for a deeply bound $K^-pp$ state, following the predictions in \cite{Yamazaki:2007hj,Yamazaki:2007cs}, that it is produced in $p$-$p$ reactions with high probability at large momentum transfer. The mass $M_X$ from (\ref{eq:X}) is close to the mass $M_{\Lambda p} \sim 2.256~ GeV/c^2$ of the $K^-pp$ candidate reported in the stopped-$K^-$ experiment by FINUDA \cite{Agnello:2005qj}, and would correspond to a binding energy $B_K = (105 \pm 2)~MeV$ for $X \equiv K^-pp$. The observed $B_K$ is larger than that originally predicted by Akaishi and Yamazaki, 
and is compared in the right frame of Fig.  \ref{fig:fits} with the spectral profiles calculated for the original  (A) \cite{Akaishi:2002bg} and for two differently enhanced hypotheses (B, C) \cite{Yamazaki:2007cs} of the $\bar{K}N$ strength. 
According to \cite{Yamazaki:2007hj,Yamazaki:2007cs} a large formation of $K^-pp$ is expected in $p$-$p$ reactions, its production rate estimated to be as much as the $\Lambda_{(1405)}$ production rate ($\sim20\%$ of the total $\Lambda$ production rate). Such a large formation probability
indicates the production of a compact system
within 
less than $1.8~fm$, much shorter hence than the average $N$-$N$ distance in ordinary nuclei ($\sim 2.2~fm$). If assuming a two-body mechanism $pp\rightarrow K^+X$, no candidate other than $K^-pp$ for a $X$ with such a large formation is predicted to date.

The $\Delta M_{K^+}$ missing-mass spectrum is compared in Fig. \ref{fig:fits} with the most relevant particle-emission thresholds: $M_{K^- pp} = 2.370~GeV/c^2$, $M_{\Lambda_{(1405)} p} = 2.345~GeV/c^2$, and $M_{\Sigma^+ \pi^- p} = 2.267$ $GeV/c^2$. The present observation is limited to the non-pionic $p\Lambda$ decay mode and hence, as the observed peak is close to the $\Sigma\pi$ emission threshold, its spectral shape is expected to be unaffected by this threshold, since the pionic decay mode is suppressed to large extent. A $K^-pp$ candidate showing a $\Gamma \sim 0.12~GeV$ in the $p\Lambda$ decay mode is thus compatible with a predominance of the $\Gamma_{p\Lambda}$ partial width (as predicted by \cite{Sekihara:2009yk,ivanov2009tbp}) and points toward a $\Gamma_{p\Sigma^+ \pi^+}$ partial width much smaller that the original prediction ($\sim 60 ~MeV$, \cite{Akaishi:2008264}). 

The large observed binding energy suggest additional effects to be considered \cite{Yamazaki:2007cs,Wycech:2008wf}, and requires theoretical studies of the decay shape \cite{Akaishi:2008264} and branch \cite{Sekihara:2009yk,ivanov2009tbp} of $K^-pp$. It points toward a strong binding regime, and does not seem compatible with a shallow $\bar{K}$ binding \cite{Jido:2003cb,Magas:2005vu,Hyodo:2007jq,Dote:2008hw}.

\paragraph{\bf Acknowledgements} We are beholden to Professor Y. Akaishi for his helpful interactions with the authors. This research was partly supported
by the DFG cluster of excellence "Origin and Structure of the Universe" of Technische Universit\"at M\"unchen and by Grant-in-Aid for Scientiﬁc Research of MonbuKagakusho of Japan. One of us (T.Y.) acknowledges the support by an Award of the Alexander von Humboldt Foundation, Germany.



\end{document}